\documentclass{jaa}

\makeatletter
\let\year\relax
\newcount\year
\makeatother

\usepackage{natbib}
\usepackage{graphicx}
\usepackage{dcolumn}

\usepackage{amsmath}
\usepackage{amssymb}
\usepackage{mathtools}
\usepackage{siunitx}

\usepackage[T1]{fontenc}
\usepackage{newpxtext}
\usepackage{euler}

\usepackage[colorlinks=true, linkcolor=blue, urlcolor=magenta, citecolor=blue]{hyperref}
\usepackage{aasmacros}
\usepackage{orcidlink}

\defcitealias{2017MNRAS.472.3177M}{M17}

\usepackage{xspace}
\def\Sref#1{Sec.~\ref{#1}\xspace}
\def\Fref#1{Fig.~\ref{#1}\xspace}
\def\Eref#1{Eq.~\eqref{#1}\xspace}

\newcommand{\LensFactory}{\href{https://akmeena766.github.io/LensFactory.jl/stable/}{\texttt{LensFactory.jl}}\xspace}
\newcommand{\Examples}{\href{https://github.com/akmeena766/LensFactory_Examples}{LensFactory$\_$Examples}\xspace}
\newcommand{\Lenses}{\href{https://akmeena766.github.io/LensFactory.jl/stable/Lenses/}{\texttt{Lenses}}\xspace}
\newcommand{\MultiPlane}{\href{https://akmeena766.github.io/LensFactory.jl/stable/MultiPlane/}{\texttt{MultiPlane}}\xspace}
\newcommand{\LensModel}{\href{https://akmeena766.github.io/LensFactory.jl/stable/LensModel/}{\texttt{LensModel}}\xspace}
\newcommand{\SingularityMap}{\href{https://akmeena766.github.io/LensFactory.jl/stable/SingularityMap/}{\texttt{SingularityMap}}\xspace}

\newcommand{\smacs}{SMACS~J0723\xspace}

\begin{document}\sloppy
	
	\title{\texttt{LensFactory.jl}: A general-purpose strong lens modeling package}
	
	\author{Ashish Kumar Meena\orcidlink{0000-0002-7876-4321}}
	\affilOne{Department of Physics, Indian Institute of Science, Bengaluru 560012, India}
	
	\twocolumn[{	
		\maketitle
		\corres{akm@iisc.ac.in}
		\msinfo{xx yy zzzz}{xx yy zzzz}
		
		\begin{abstract}
			We present \LensFactory, an efficient, open-source, general-purpose strong lens modeling package written in \texttt{Julia}. It supports single- and multi-plane lensing with a wide range of analytic lens profiles, and its modular design, built on \texttt{Julia}'s multiple dispatch, makes it straightforward to add new profiles or inference methods, while native multi-threading parallelizes the computationally expensive steps, enabling full cluster-scale lens reconstructions within hours on a desktop-class machine. We validate the mass-reconstruction pipeline of \LensFactory on two simulated galaxy clusters (Ares and Hera), recovering the convergence and magnification maps well where the lensing constraints lie, and on the JWST-observed cluster SMACS~J0723.3$-$7327, obtaining a strong lens model (image-plane RMS of $0.46''$) consistent with previously published results. In addition, we introduce singularity maps as a complementary diagnostic for comparing lens models, probing differences in the higher-order derivatives of the lensing potential, which remain largely unconstrained by standard observables such as image positions, time delays, and flux ratios. \LensFactory is accompanied by a set of worked examples covering all core functionality.    
		\end{abstract}
		
	}]

	\section{Introduction}
	\label{sec:intro}
	Gravitational lensing refers to the bending of light as it passes close to an intervening mass distribution~(referred to as a \emph{lens}) between the distant source and the observer~\citep[e.g.,][]{1992grle.book.....S, 1996astro.ph..6001N, 2001stgl.book.....P}. \emph{Strong lensing} occurs when the lens is sufficiently dense to produce multiple images of a background source. Since lensing is independent of the dynamical state of the lens, one of the primary uses of strong lensing is to reconstruct the (visible and dark) lens mass distribution\footnote{We note that weak lensing is also used for lens mass reconstruction~\citep[e.g.,][]{2001PhR...340..291B}. However, here we focus only on strong lensing.}. The reconstructed lens models are then used to infer properties of the Universe and its components. For example, the time delay between multiple images is used to measure the Hubble constant~\citep[$H_0$;][]{2020MNRAS.498.1420W, 2023Sci...380.1322K, 2025ApJ...979...13P, 2026A&A...708A.291S}. The magnification estimates are essential for studying properties of parsec-scale structures~\citep[e.g.,][]{2024Natur.632..513A, 2025arXiv251208054A, 2026arXiv260110097M} and individual stars~\citep[e.g.,][]{2018NatAs...2..334K, 2022Natur.603..815W, 2023MNRAS.521.5224M, 2023ApJ...944L...6M} in distant sources all the way into the first billion years of the Universe. Modeling anomalous features in observed lensed images allow us to constrain the nature of the elusive dark matter~\citep[e.g.,][]{2020MNRAS.491.6077G, 2024SSRv..220...58V, 2026arXiv260605277G}.
	
	Lens mass reconstruction via strong lensing is an ill-posed problem, as only a finite number of lensed images (constraints) are available, providing only local constraints. Hence, one needs to make additional (physically motivated) assumptions about the lens mass distribution, thereby narrowing the range of allowed parameter values. Overall, the lens modeling methods can be broadly divided into two categories: parametric and free-form. Parametric methods describe the lens mass distribution using analytical profiles (e.g., Navarro-Frenk-White or isothermal ellipsoids), typically associating mass components with the luminous cluster galaxies~\citep[e.g.,][]{2005ApJ...621...53B, 2007NJPh....9..447J, 2015ApJ...801...44Z}. On the other hand, free-form (or non-parametric) methods do not assume any specific shape for the lens mass profile. Instead, they describe the lens using pixels or basis functions~\citep[e.g.,][]{1998MNRAS.294..734A, 2005A&A...437...39B, 2005MNRAS.360..477D, 2007MNRAS.380.1729L}; although such decompositions are themselves parametrized, the parameters carry no assumption about the form of the underlying mass distribution, and the solution is driven by the lensing observables. While free-form methods offer greater flexibility and avoid certain inherent biases of the parametric methods, they are severely under-constrained and rely on regularization techniques to prevent the introduction of spurious mass components.
	
	Even within parametric lens modeling methods, the reconstructed lens mass distribution can differ considerably due to the underlying model assumptions. For example, \citet{2017MNRAS.472.3177M} showed that even when different modelers used the same tool to model the same cluster, the best-fit lens models can show substantial variation; the differences can be much more significant across free-form methods. A similar spread is evident in recent measurements of the Hubble constant~($H_0$) from cluster lensing~\citep{2025ApJ...979...13P, 2026A&A...708A.291S}, where the~$H_0$ inferred from individual lens models, even after weighting by their time-delay predictions, varies enough to be consistent with both early- and late-time $H_0$ measurements. Understanding and quantifying such model-to-model variation requires accessible, extensible modeling tools as well as new diagnostics for comparing lens models beyond the standard observables. In particular, modeling of the same lens with independently developed codes complements same-tool comparisons. While the latter isolate modeler-driven choices~(e.g., priors, number of mass components, selection of image systems), only the former can expose implementation-specific systematics, which remain perfectly correlated among analyses sharing a single code. In addition, agreement among independent codes strengthens confidence in the inferred lens, source, and cosmological properties. These considerations form the primary motivation behind the present work.
	
	In this work, we present \LensFactory\footnote{\url{https://github.com/akmeena766/LensFactory.jl}}, an open-source, general-purpose strong lens modeling package. Several mature packages already exist for strong lens modeling --- e.g., 
	\texttt{Lenstool}~\citep{1993A&A...273..367K, 2007NJPh....9..447J}, 
	\texttt{glafic}~\citep{2010PASJ...62.1017O}, 
	\texttt{lenstronomy}~\citep{2018PDU....22..189B}, and 
	\texttt{PyAutoLens}~\citep{2021JOSS....6.2825N} --- written primarily in C or Python\footnote{More recently, \texttt{Gravity.jl}~\citep{2024A&A...690A.346L}, a \texttt{Julia}-based lens modeling framework has also been presented; however, the code is proprietary and not publicly available at the time of writing.}.
	\LensFactory complements these efforts as an independently developed, open-source package written in \texttt{Julia}~\citep{Julia2017}, a high-performance, dynamic programming language. This choice of language rests on three factors. First, \texttt{Julia}'s LLVM-based just-in-time compilation gives near-C performance for numerically intensive calculations without a separate low-level backend, avoiding the two-language problem while keeping the code readable. Second, multiple dispatch lets us define generic interfaces that automatically specialize to different lens mass profiles and inference methods, making it easy to add new profiles or samplers. Third, native multi-threading parallelizes expensive tasks such as posterior sampling across chains or walkers with minimal overhead. \texttt{Julia}'s package extension mechanism also keeps optional dependencies (e.g., plotting) out of the core, enabling a lightweight installation.
	
	The current manuscript is organized as follows. In \Sref{sec:basics}, the basics of single- and multi-plane gravitational lensing are briefly discussed. \Sref{sec:model} discusses the relevant basics of lens modeling and its implementation in \LensFactory. In \Sref{sec:sims} and \Sref{sec:smacs}, we present \LensFactory reconstructed lens models for simulated galaxy clusters~(Ares and Hera) from~\citet{2017MNRAS.472.3177M} and SMACS~J0723.3$–$7327, respectively. Singularity maps, as an additional diagnostic to compare lens models, are discussed in \Sref{sec:singularity}. Summary and future prospects are presented in \Sref{sec:summary}. Unless mentioned otherwise, we use $(H_0, \Omega_{m,0}, \Omega_{w,0}) = (70~\mathrm{km/s/Mpc}, 0.3, 0.7)$, where $\Omega_{w,0}$ denotes the dark-energy density parameter with equation of state $w$~($=-1$; unless stated otherwise). All magnitudes are given in the AB system~\citep{1983ApJ...266..713O}.

	\section{Basics}
	\label{sec:basics}
	
	\subsection{Single-plane lensing}
	\label{ssec:single}
	Arguably, one of the most fundamental concepts in gravitational lensing is the Fermat potential~($\phi$) or arrival time delay function ($t_d$), given as~\citep[e.g.,][]{1986ApJ...310..568B, 1992grle.book.....S, 2001stgl.book.....P},
	\begin{equation}
		\begin{split}
			t_d(\pmb{\theta}, \pmb{\beta}) 
			&= \frac{1+z_d}{\rm{c}} \frac{D_d D_s}{D_{ds}} \theta_0^2 \: \phi(\pmb{\theta}, \pmb{\beta}) \\
			=& \frac{1+z_d}{\rm{c}} \frac{D_d D_s}{D_{ds}} \theta_0^2
			\left[ \frac{(\pmb{\theta} - \pmb{\beta})^2}{2} - \frac{D_{ds}}{D_s} \psi(\pmb{\theta}) \right],
		\end{split}
		\label{eq:time_delay}
	\end{equation}
	where $z_d$ is the lens redshift. $D_d$, $D_{ds}$, and $D_s$ are angular diameter distances from observer to lens, lens to source, and observer to source, respectively. $\theta_0$ is an (arbitrary) angular scale (set to $1''$ in \LensFactory). $\pmb{\theta}$ and $\pmb{\beta}$ are (normalized) 2D angular vectors in lens and source plane, respectively.  $\psi(\pmb{\theta})$ represents the projected lensing potential and given as,
	\begin{equation}
		\psi(\pmb{\theta}) = \frac{4\,\rm{G}}{\rm{c^2}} \frac{1}{D_d} \int d^2\pmb\theta' \Sigma(\pmb{\theta}') \ln|\pmb{\theta} - \pmb{\theta}'|,
	\end{equation}
	where~$\Sigma(\pmb{\theta})$ is the projected surface mass density~(in angular units) in the lens plane. We note that, unlike the conventional definition used in gravitational lensing, the projected lensing potential is defined here without the distance ratio,~$a_{\rm dis} \coloneqq D_{ds}/D_s$. This choice is convenient because, in single-plane lensing, source-redshift dependence enters lensing quantities only through this factor. 
	
	For a given source position, the stationary points of the Fermat potential mark the locations of lensed images in the image plane. Equating the gradient of the Fermat potential to zero gives the \emph{lens equation}, a mapping between the image plane and the source plane, which can be written as,
	\begin{equation}
		\pmb{\beta} = \pmb{\theta} - a_{\rm dis} \pmb{\alpha}(\pmb{\theta}),
		\label{eq:lens_eq}
	\end{equation}
	where $\pmb{\alpha}=\pmb{\nabla}\psi$ is the deflection vector. The lens mapping is non-linear. This (often) leads to the formation of multiple images for a given source position, i.e., strong lensing. The properties of the observed lensed images can be described by the Jacobian matrix of the lens equation, written as,
	\begin{equation}
		\mathbb{A}(\pmb{\theta}) \coloneqq \frac{\partial\pmb{\beta}}{\partial\pmb{\theta}} \quad 
		\Rightarrow \quad \mathbb{A}_{ij} = \delta_{ij} - a_{\rm dis} \psi_{ij}
		\label{eq:jac}
	\end{equation}
	where subscripts represent the partial derivatives~(i.e.,~$\psi_{ij}=\partial^2\psi/\partial\theta_i\partial\theta_j$) and~$\psi_{ij}$ is known as the \emph{deformation tensor}, which describes the distortions in the observed lensed images. It is given as,
	\begin{equation}
		\psi_{ij} = 
		\begin{pmatrix}
			\psi_{11} & \psi_{12} \\
			\psi_{21} & \psi_{22}
		\end{pmatrix}
		=
		\begin{pmatrix}
			\kappa + \gamma_1 & \gamma_2 \\
			\gamma_2 & \kappa - \gamma_1
		\end{pmatrix}.
	\end{equation}
	In the above equation, we have introduced the well-known convergence~($\kappa$) and the components of the shear tensor~($\gamma \coloneqq \gamma_1 + i \gamma_2$), which can be written in terms of derivatives of projected lensing potential as,
	\begin{equation}
		\begin{split}
			\kappa   &= \frac{1}{2}(\psi_{11} + \psi_{22}), \\
			\gamma_1 &= \frac{1}{2}(\psi_{11} - \psi_{22}), \\ 
			\gamma_2 &= \psi_{12}.
		\end{split}
	\end{equation}
	The convergence~($\kappa$), as it depends on the trace of the Jacobian matrix, governs the isotropic distortion in the observed image. On the other hand, the shear (components) governs the directional distortion of the observed image. We note that the above defined~$(\kappa, \gamma)$ values differ from the canonical definition by~$a_{\rm dis}$ factor. The magnification factor of a lensed image (corresponding to a point source) formed at~$\pmb{\theta}$ is defined as,
	\begin{equation}
		\mu(\pmb{\theta}) \coloneqq \frac{1}{\rm{det}\mathbb{A}} 
		= \frac{1}{(1-a_{\rm dis} e_1) (1-a_{\rm dis} e_2)},
	\end{equation}
	where $e_1\coloneqq\kappa+|\gamma|$ and $e_2\coloneqq\kappa-|\gamma|$ represent the eigenvalues of the deformation tensor. We note that for certain combinations of ($\kappa, |\gamma|$)~values in the image plane, the magnification can diverge~(i.e.,~$\mu^{-1}=0$). Such points form smooth closed curves in the image plane, known as \emph{critical curves}. The corresponding (not necessarily smooth) curves in the source plane are known as \emph{caustics}. The infinite magnification arises from the point source approximation above. However, for a finite-size source, we take a weighted average over the source, yielding finite magnification values\footnote{Even for a point source, wave optics would lead to finite magnification values.}. In a typical cluster lens modeling scenario, one identifies counter-images ``knots''~(point-like features typically associated with star-forming regions and star clusters) in the observed lensed images, and the above magnification formula is enough for such cases. 
	
	All of the lensing quantities described above are implemented in \LensFactory. The package currently has more than 20 lens models available, and thanks to its highly modular nature, it is straightforward to add new ones. The single-plane lensing~(by an isolated or complex lens model) is controlled by the \Lenses module, which has functions to calculate essentially all standard lensing quantities. For elliptical lenses, the ellipticity is defined as~$\epsilon\coloneqq(1-q)/(1+q)$, where~$q$ is the axis ratio. In addition, the position angle~($\phi_{\rm PA}$) is measured with respect to the x-axis in a counter-clockwise direction. For examples on the use of various lens models in single-plane lensing, we refer readers to the \Examples repository.

	\subsection{Multi-plane lensing}
	\label{ssec:multi}
	Often, in strong lensing (especially by galaxy clusters), we encounter line-of-sight~(LOS) structures that also influence the observed lensing features. Neglecting such structures and relying solely on a single-plane approximation can introduce systematic biases in the inferred lens, source, and cosmological properties. In such cases, we need to account for multiple successive deflections of a light ray~\citep[determined by the number of lens planes; see chapter 9 in][]{1992grle.book.....S} before it reaches the observer. Here we briefly revisit the basic formulae and refer readers to \citet{1992grle.book.....S} and \citet{2001stgl.book.....P} for more details.
	
	In a general case of lensing by N-lens plane, the lens equation (after incorporating deflection in all lens planes) can be written as,
	\begin{equation}
		\pmb{\beta} = \pmb{\theta}_1 - \sum_{j=1}^N \frac{D_{js}}{D_s} \pmb{\alpha}_j(\pmb{\theta}_j),
	\end{equation}
	where~$D_{js}$ and $D_s$ are the angular diameter distances from the $j$-th lens plane to the source and from the observer to the source, respectively. $\pmb{\alpha}_j(\pmb{\theta}_j)$ represents the deflection in $j$-th lens plane at position~$\pmb{\theta}_j$. Since the light ray is deflected in each plane, the impact parameter in $j$-th plane would be given as,
	\begin{equation}
		\pmb{\theta}_j = \pmb{\theta}_1 - \sum_{i=1}^{j-1} \frac{D_{ij}}{D_j} \pmb{\alpha}_i(\pmb{\theta}_i),
	\end{equation}
	where $D_{ij}$ and $D_j$ are the angular diameter distances between $i$-th to $j$-th lens planes and from observer to the $j$-th plane, respectively. Similar to single-plane lensing, we can again define lensing potential for $j$-th lens plane as,
	\begin{equation}
		\psi_j(\pmb{\theta}_j) = \frac{4\,\rm{G}}{\rm{c^2}} \frac{1}{D_j} \int d^2\pmb{\theta}' \Sigma_j(\pmb{\theta}') \ln|\pmb{\theta}_j - \pmb{\theta}'|,
	\end{equation}
	where~$\Sigma_j(\pmb{\theta}_j)$ is the projected surface mass density in the $j$-th lens plane at position~$\pmb{\theta}_j$. With that, the total lensing potential and deflection angle (summed over all lens planes) can be written as,
	\begin{align}
		\psi(\pmb{\theta}_1, ..., \pmb{\theta}_N)   &= \sum_{j=1}^N \frac{D_{js}}{D_s} \psi_j (\pmb{\theta}_j), \\
		\alpha(\pmb{\theta}_1, ..., \pmb{\theta}_N) &= \sum_{j=1}^N \frac{D_{js}}{D_s} \alpha_j (\pmb{\theta}_j).
	\end{align}
	Again, we note that the definitions of various lensing quantities differ from the typical definitions presented in~\citet{1992grle.book.....S} in terms of distance ratios. Finally, the time delay function for multi-plane lensing is given as,
	\begin{equation}
		T(\pmb{\theta}_1, ..., \pmb{\theta}_N, \pmb{\beta}) = \sum_{j=1}^N T_{j, j+1}(\pmb{\theta}_j, \pmb{\theta}_{j+1}); \quad \pmb{\theta}_{N+1} = \pmb{\beta}
	\end{equation}
	with
	\begin{multline}
		T_{j, j+1}(\pmb{\theta}_j, \pmb{\theta}_{j+1}) = \\ \frac{1+z_j}{\rm c} \frac{D_j D_{j+1}}{D_{j,j+1}} \theta_0^2 \left[\frac{(\pmb{\theta}_j - \pmb{\theta}_{j+1})^2}{2} - \frac{D_{j,j+1}}{D_{j+1}}\psi_j(\pmb{\theta}_j)\right],
	\end{multline}
	where~$z_j$ is the redshift of $j$-th lens plane. The two terms in brackets represent the geometric and gravitational contributions to the time delay in each lens plane, respectively. In multi-plane lensing, we can again define the magnification similar to the single-plane case, i.e.,~$\mu\coloneqq1/det\mathbb{A}$. However, the lens mapping is no longer a gradient mapping and the corresponding Jacobian matrix is not symmetric~\citep[e.g.,][]{1992grle.book.....S, 2001stgl.book.....P}, which introduces a net rotation in the lensed images. 
	
	In \LensFactory, multi-plane lensing is controlled by the \MultiPlane module, which serves as the entry point for all related functions and mirrors the interface of the \Lenses module for single-plane lensing. For related examples, we again refer readers to the \Examples repository.

	\subsection{Solving lens equation}
	\label{ssec:leq_solve}
	Predicting the positions of lensed images is a central task in gravitational lensing; it is the primary observable that constrains the lens mass distribution, yet it requires inverting the lens equation, which is, in general, non-linear and multi-valued. At present, \LensFactory solves the lens equation numerically by recasting it as a root-finding problem. Specifically, defining $\pmb{f}(\pmb{\theta})\coloneqq\pmb{\beta}-\pmb{\theta}+\pmb{\alpha}$, the image positions correspond to the zeros of~$f$, i.e., where~$f_1(\pmb{\theta})=0=f_2(\pmb{\theta})$. The iso-contours of each component are computed over a user-defined image-plane grid using the marching squares algorithm, and the image positions are identified as their intersection points. These intersections are determined by solving for the exact crossing of each pair of piecewise-linear contour segments, yielding the lensed image positions.
	
	The accuracy of the recovered image positions scales as~$\delta \theta \sim \mathcal{O}(\delta h^2)$ for typical source positions, where~$\delta h$ is the grid pixel size. Since the segment intersections are computed exactly, the error arises solely from the piecewise-linear approximation of the underlying smooth iso-contours, which carries a second-order error in~$\delta h$. For images very close to critical curves, however, the two contours can become nearly parallel to each other, and the positional error can be substantially larger, though this only becomes relevant when the source lies extremely close to a caustic.

	\begin{figure*}[!h]
		\centering
		\includegraphics[width=1.0\linewidth]{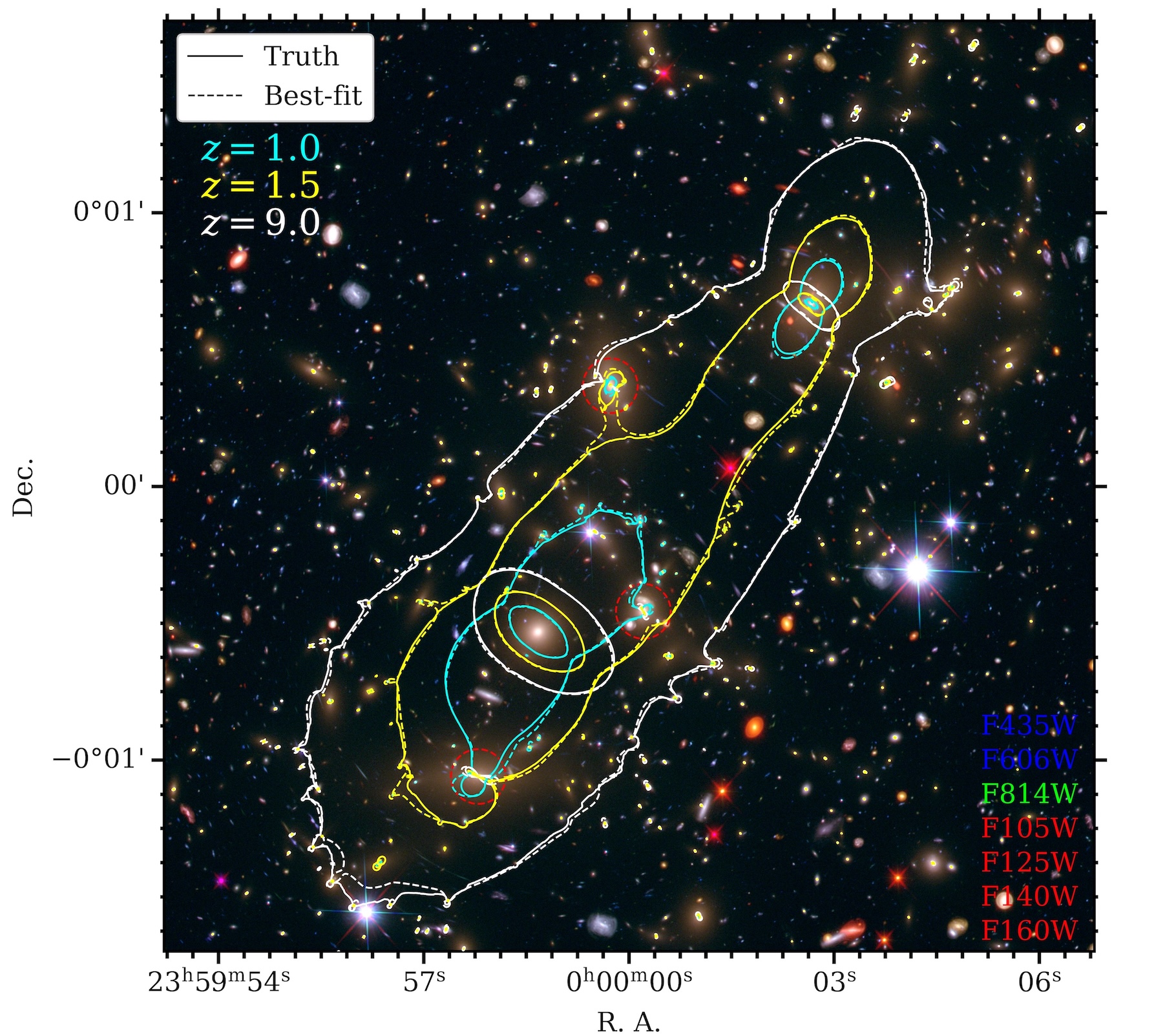}
		\caption{Color image of simulated galaxy cluster Ares~($z=0.5$). The solid and dashed curves represent critical curves at $z=1.0,~1.5,~9.0$ corresponding to actual~(i.e., true) and reconstructed mass distribution, respectively. The red-dashed circles mark the position of three galaxies, which are modeled individually~(i.e., outside the scaling relations).}
		\label{fig:ares}
	\end{figure*}
	
	\begin{figure*}[!h]
		\centering
		\includegraphics[width=1.0\linewidth]{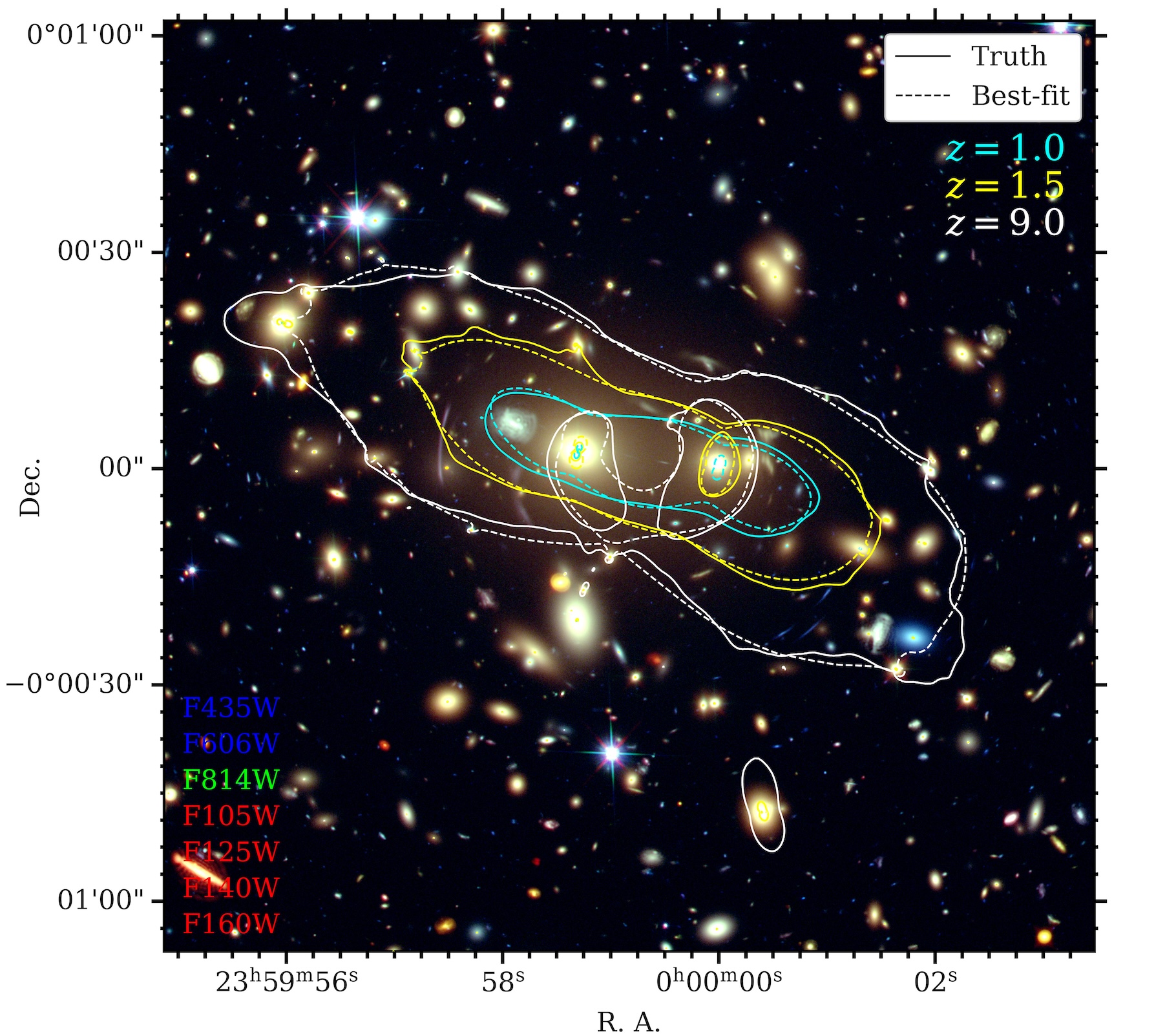}
		\caption{Color image of simulated galaxy cluster Hera~($z=0.507$). The solid and dashed curves represent critical curves at $z=1.0,~1.5,~9.0$ corresponding to actual~(i.e., true) and reconstructed mass distribution, respectively.}
		\label{fig:hera}
	\end{figure*}
	
	\section{Lens modeling}
	\label{sec:model}
	In parametric lens modeling, the lens mass distribution is described by a set of analytic components, and the corresponding free parameters are constrained by requiring the model to reproduce the observed lensing quantities. The goodness of fit is quantified
	through a total $\chi^2 = \chi^2_{\rm img} + \chi^2_{\rm f} + \chi^2_{\Delta t}$, whose terms correspond to the image positions, fluxes, and time delays, with the associated likelihood~$\mathcal{L} \propto \exp(-\chi^2/2)$. Below, we briefly describe each term, largely following \citet{2001astro.ph..2340K} and \citet{2010PASJ...62.1017O}, along with its source-plane approximation, followed by the treatment of cluster member galaxies. The best-fit model is then obtained by minimizing the total~$\chi^2$ (equivalently, maximizing the likelihood), and the parameter uncertainties are estimated by sampling the corresponding posterior. Below, we describe various~$\chi^2$ terms and their approximate source plane version.
	
	\subsection{Image position}
	\label{ssec:chi2_position}
	The positions of the multiple images are the primary constraints in strong lens modeling. Assuming independent Gaussian astrometric uncertainties, the image-plane $\chi^2$ is given as,
	\begin{equation}
		\chi_{\rm img}^2 = \sum_{i} \delta \pmb{\theta}_i^T \cdot S_i^{-1} \cdot \delta \pmb{\theta}_i,
		\label{eq:chi2_img}
	\end{equation}
	where,
	\begin{equation}
		\delta \pmb{\theta}_i = \pmb{\theta}_{i,\rm{obs}} - \pmb{\theta}_{i,\rm{model}},
	\end{equation}
	is the offset between the observed ($\boldsymbol{\theta}_{\rm obs}$) and model-predicted ($\boldsymbol{\theta}_{\rm model}$) image position. The $S_i$ represents the position error covariance matrix for the $i$-th image, and given as,
	\begin{equation}
		S_i = R_i^T 
		\begin{bmatrix}
			\sigma_{1,i}^2 & 0 \\
			0 & \sigma_{2,i}^2
		\end{bmatrix}
		R_i,
	\end{equation}
	where
	\begin{equation}
		R_i = 
		\begin{bmatrix}
			\cos\phi_{\sigma,i} & \sin\phi_{\sigma,i} \\
			-\sin\phi_{\sigma,i} & \cos\phi_{\sigma,i}
		\end{bmatrix}.
	\end{equation}
	In the above, $\sigma_{1,i}$ and $\sigma_{2,i}$ are semi-major and semi-minor axes of the error ellipse, and $\phi_{\sigma,i}$ is its position angle measured with respect to the $x$-axis in the counter-clockwise direction; for circular uncertainties, $S_i$ reduces to $\sigma_i^2$ times the identity matrix.
	
	Evaluating \Eref{eq:chi2_img} requires solving the (non-linear) lens equation for every source at every step of the minimization to obtain the model-predicted image positions. This makes image-plane minimization computationally expensive, particularly for cluster lenses with tens of sources. Instead of $\chi_{img}^2$ minimization in the image plane, one can evaluate an approximate $\chi_{src}^2$ in the source plane, which avoids solving the lens equation altogether~\citep[e.g.,][]{2001astro.ph..2340K}, as,
	\begin{equation}
		\chi_{\rm src}^2 = \sum_i \delta \pmb{\beta}_i^T \cdot \mu_i^T \cdot S_i^{-1} \cdot \mu_i \cdot \delta \pmb{\beta}_i,
		\label{eq:chi2_src}
	\end{equation}
	where 
	\begin{equation}
		\delta \pmb{\beta}_i \coloneqq \pmb{\beta}_{i,\rm{obs}} - \pmb{\beta}_{\rm{model}},
	\end{equation}
	is the source plane offset between the source position obtained by mapping the $i$-th observed image to the source plane($\boldsymbol{\beta}_{i,{\rm obs}}$), and the (common) model source position~(~$\boldsymbol{\beta}_{\rm model}$). The~$\mu_i=\mathbb{A}^{-1}$ is the magnification tensor (inverse Jacobian) evaluated at the observed image position. The~$\chi^2_{src}$ is an approximate version of~$\chi^2_{img}$ assuming that the separation between observed and predicted image positions is small~
	\begin{equation}
		\pmb{\theta}_{i,\rm{obs}} - \pmb{\theta}_{i, \rm{model}} \approx \mathbb{A}^{-1} (\pmb{\beta}_{i,\rm{obs}} - \pmb{\beta}_{\rm{model}}).
		\label{eq:delta}
	\end{equation}
	Retaining the full tensor, rather than a scalar magnification correction, is essential for it to faithfully track its image-plane counterpart~\citep{2010PASJ...62.1017O}. 
	
	Because the true unlensed source position~$\pmb{\beta}_{\rm{model}}$ is not observable, it can be treated as a nuisance parameter and determined analytically via linear minimization (i.e.,
	$\partial \chi^2_{\rm src}/\partial \pmb{\beta}_{\rm{model}} = 0$),
	\begin{equation}
		\pmb{\beta}_{\rm model} = G^{-1} \cdot \pmb{g},
	\end{equation}
	where,
	\begin{equation}
		\begin{aligned}
			G =& \sum_i \mu_i^T \cdot S_i^{-1} \cdot \mu_i, \\
			\pmb{g} =& \sum_i \mu_i^T \cdot S_i^{-1} \cdot \mu_i \cdot \pmb{\beta}_{i, \rm obs}.
		\end{aligned}
	\end{equation}

	\begin{figure*}[!h]
		\centering
		\includegraphics[width=0.48\linewidth]{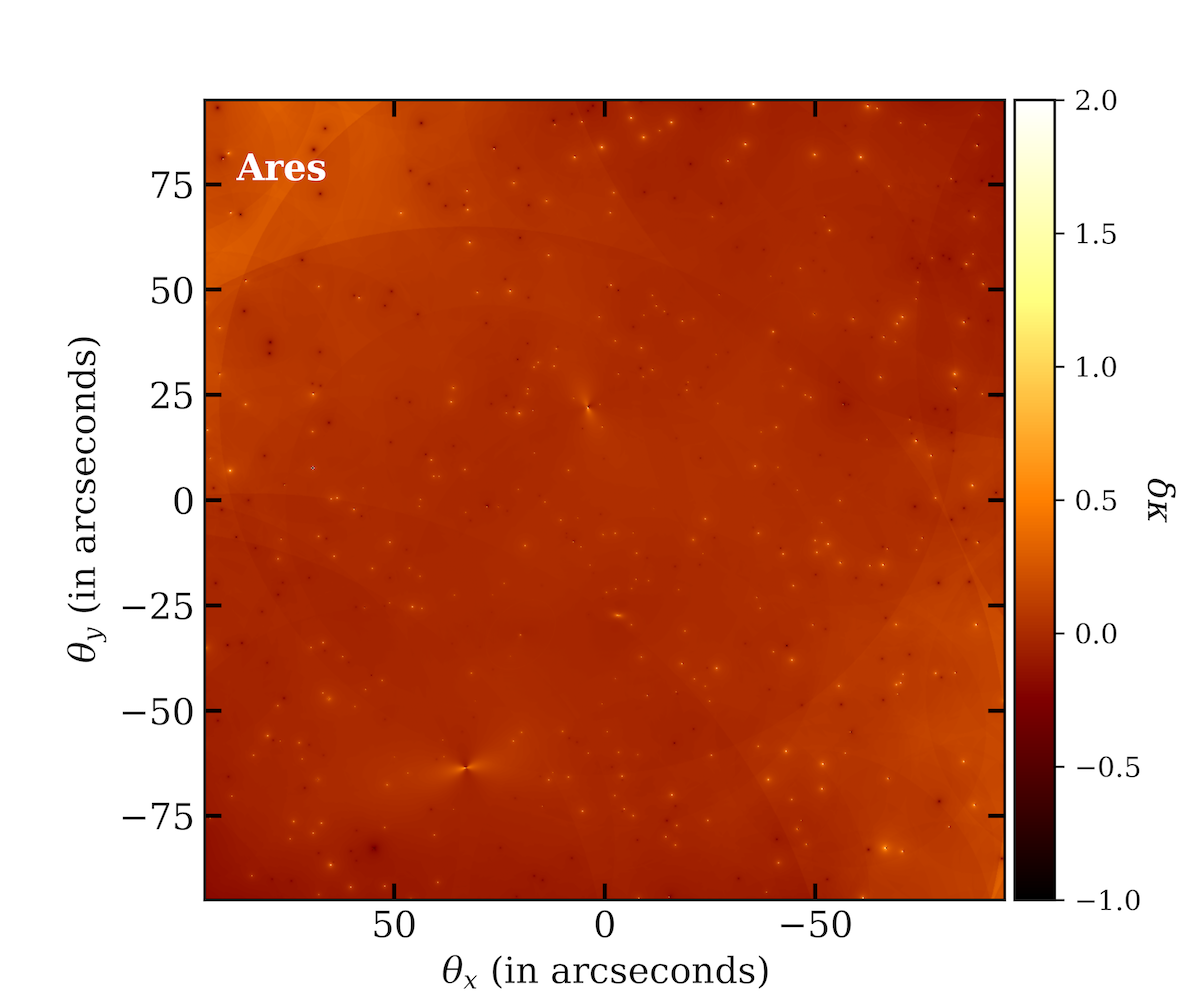}
		\includegraphics[width=0.48\linewidth]{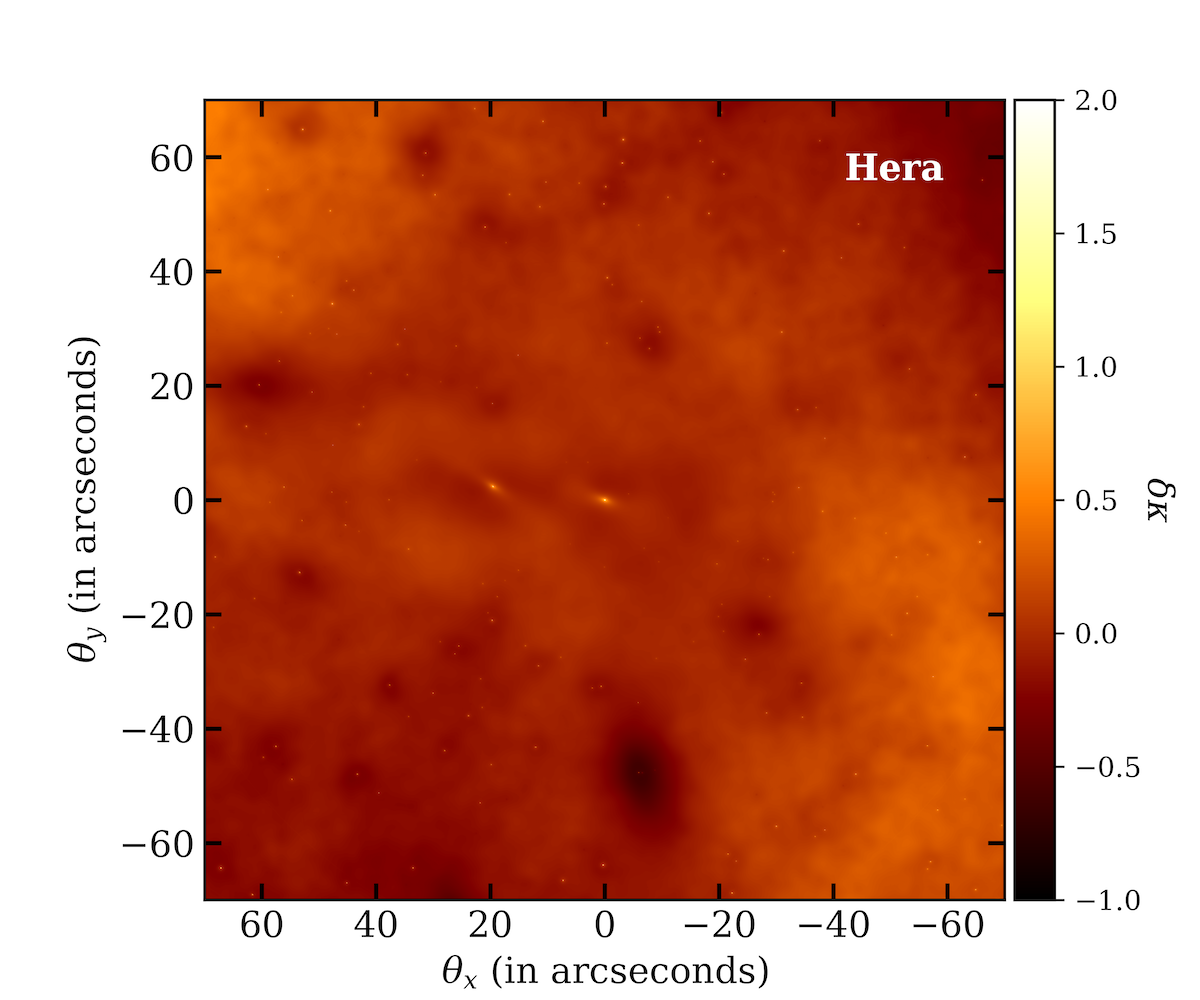}
		\caption{Relative difference between true and best-fit convergence maps,~$\delta\kappa=(\kappa_{\rm{model}} - \kappa_{\rm{true}})/\kappa_{\rm{true}}$, for a source at $z=9$. The left and right panels correspond to Ares and Hera, respectively.}
		\label{fig:dkappa}
	\end{figure*}

	\subsection{Flux}
	\label{ssec:chi2_flux}
	When reliable photometry is available --- and systematics such as microlensing, substructure, or differential extinction are under control --- the fluxes (or magnitudes) of lensed images constrain the local magnification and, hence, the second derivatives of the lens potential. Assuming that the photometric errors for different lensed images are independent and Gaussian, the~$\chi_f^2$ for observed fluxes (in observed magnitudes) is given as~\citep{2001astro.ph..2340K},
	\begin{equation}
		\chi_f^2 = \sum_i \frac{\left(m_i + 2.5\log|\mu_i| - m_{\rm src}\right)^2}{\sigma_{m_i}^2},
		\label{eq:chi2_flux}
	\end{equation}
	where~$m_i$ represents the observed magnitude of the $i$-th image and~$\sigma_{m_i}$ is the associated error. $\mu_i$ represents the magnification at the model-predicted image position~($\pmb{\theta}_{i,\rm{model}}$). The intrinsic source magnitude,~$m_{\rm src}$, remains unobserved. Similar to~$\pmb{\beta}_{\rm model}$ in \Sref{ssec:chi2_position}, it can be treated as a nuisance parameter and determined analytically by solving~$\partial\chi_f^2/\partial m_{\rm src}=0$, yielding,
	\begin{equation}
		m_{\rm src} = \frac{\sum_i\left[(m_i + 2.5\log|\mu_i|)/\sigma_{m_i}^2\right]}{\sum_i(1/\sigma_{m_i}^2)}.
		\label{eq:msrc}
	\end{equation}
	Evaluating the model magnification in \Eref{eq:chi2_flux} again requires solving the lens equation to locate the model-predicted image positions. However, assuming that the separation between the observed and model-predicted image positions is small, the magnification can be Taylor-expanded around the observed position~\citep{2010PASJ...62.1017O},
	\begin{equation}
		\begin{aligned}
			\mu_{i,\rm{model}} \approx
			& \: \mu(\pmb{\theta}_{i,\rm{obs}}) \\
			& + (\pmb{\theta}_{i,\rm{model}}-\pmb{\theta}_{i,\rm{obs}}) \cdot \left[\nabla_{\pmb{\theta}}\mu\right]_{\pmb{\theta}_{i,\rm{obs}}}.
		\end{aligned}
	\end{equation}
	Substituting the first-order position approximation, \Eref{eq:delta}, into this expression gives the linear expansion for the magnification,
	\begin{equation}
		\begin{aligned}
			\mu_{i,\rm{model}} \approx
			& \: \mu(\pmb{\theta}_{i,\rm{obs}}) \\
			& + \mathbb{A}^{-1} (\pmb{\beta}_{\rm{model}}-\pmb{\beta}_{i,\rm{obs}}) \cdot \left[\nabla_{\pmb{\theta}}\mu\right]_{\pmb{\theta}_{i,\rm{obs}}},
		\end{aligned}
	\end{equation}
	which involves only quantities evaluated at the observed image positions and the model source position, and hence carries no additional computational cost within the source-plane minimization.

	\begin{figure*}[!h]
		\centering
		\includegraphics[width=0.48\linewidth]{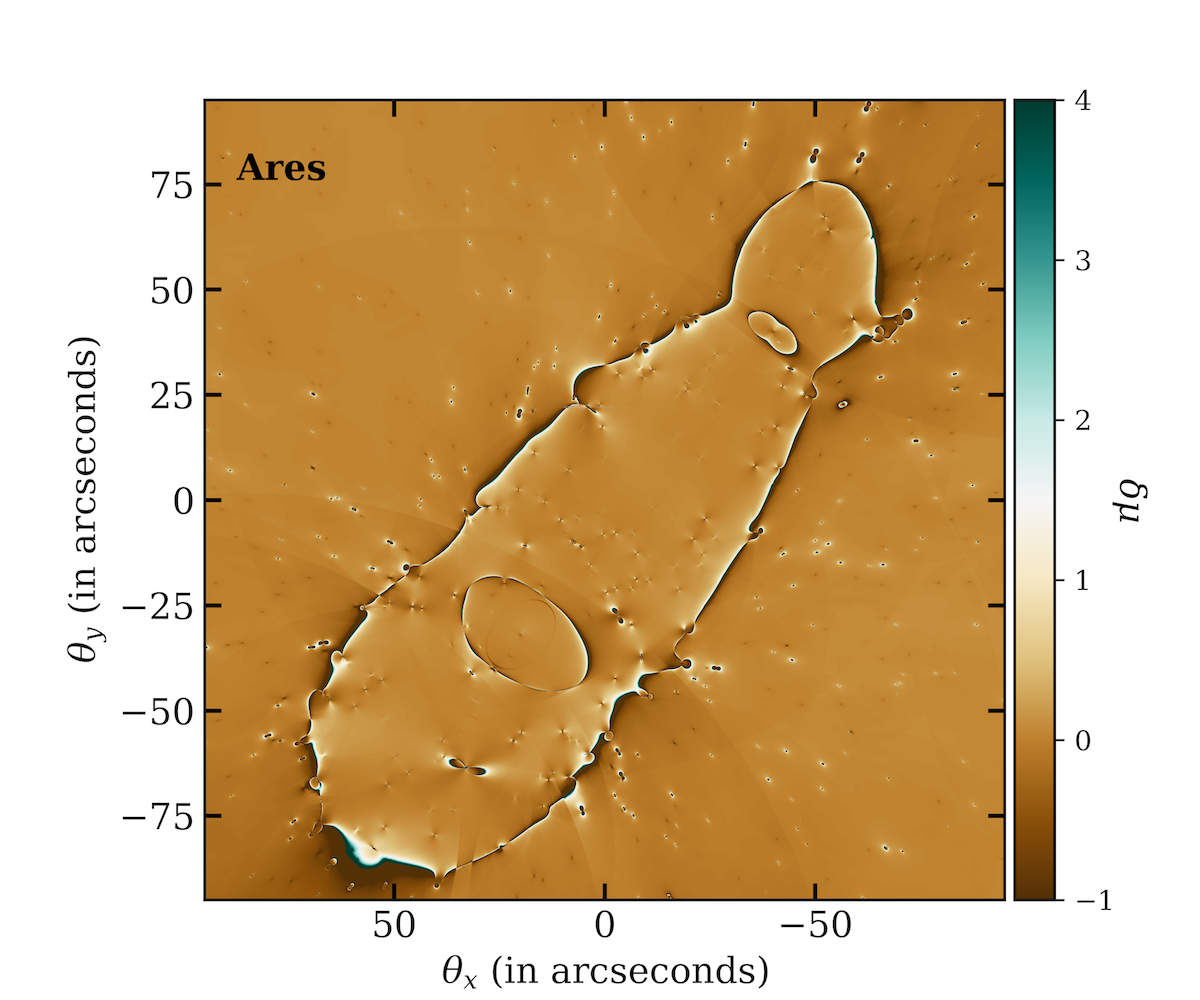}
		\includegraphics[width=0.48\linewidth]{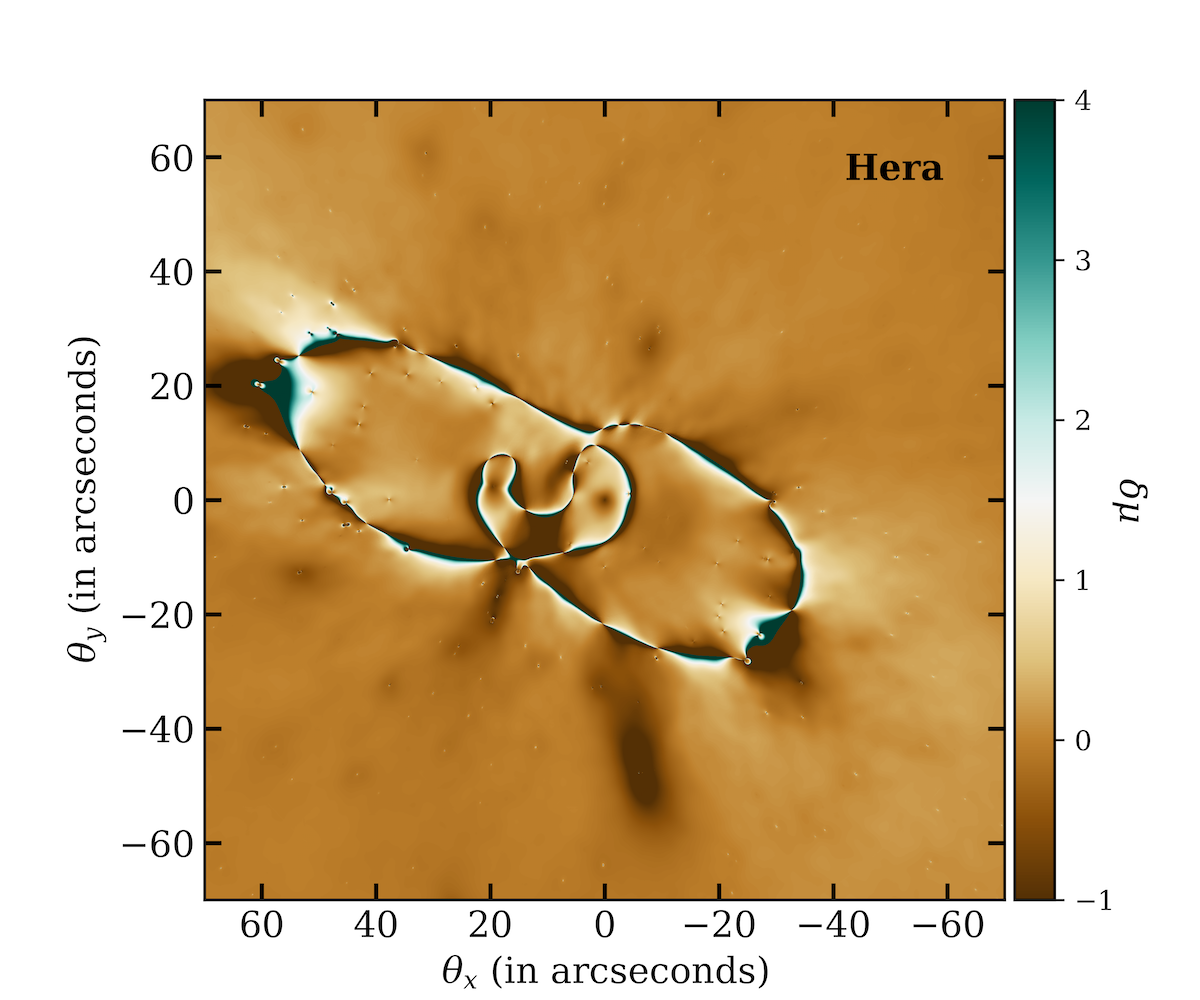}
		\caption{Relative difference between true and best-fit magnification maps,~$\delta\mu=(\mu_{\rm{model}} - \mu_{\rm{true}})/\mu_{\rm{true}}$, for a source at $z=9$. The left and right panels correspond to Ares and Hera, respectively.}
		\label{fig:dmu}
	\end{figure*}

	\subsection{Time delay}
	\label{ssec:chi2_td}
	For variable background sources (e.g., quasars and supernovae), the relative time delays between lensed images can be measured, providing constraints on the Fermat potential differences between the image positions, see \Eref{eq:time_delay}, and, through the time-delay distance, on the Hubble constant. Assuming independent Gaussian uncertainties on the measured delays, the corresponding $\chi^2$ is given as,
	\begin{equation}
		\chi_{\Delta t}^2 = \sum_i\frac{\left(\Delta t_{i,\rm{obs}} - \Delta t_{i,\rm{model}} - \Delta t_0\right)^2}{\sigma_{\Delta t_i}^2},
		\label{eq:chi2_td}
	\end{equation}
	where $\Delta t_{i,\rm{obs}}$ and $\Delta t_{i,\rm{model}}$ are the observed and model-predicted time delays of the $i$-th image, and $\sigma_{\Delta t_i}$ is the error associated with the measured delay. Since only relative delays are observable, the common zero-point, $\Delta t_0$, remains unconstrained; similar to~$\pmb{\beta}_{\rm model}$ and~$m_{\rm src}$ above, it can be treated as a nuisance parameter and determined analytically by solving~$\partial\chi_{\Delta t}^2/\partial \Delta t_0 = 0$, yielding,
	\begin{equation}
		\Delta t_0 = \frac{\sum_i\left[(\Delta t_{i,\rm{obs}} - \Delta t_{i,\rm{model}})/\sigma_{\Delta t_i}^2\right]}{\sum_i(1/\sigma_{\Delta t_i}^2)}.
		\label{eq:dt0}
	\end{equation}
	As with the magnification in \Sref{ssec:chi2_flux}, evaluating $\Delta t_{i,\rm{model}}$ exactly requires the model-predicted image positions, which are not available during source-plane minimization. We therefore expand the time delay to first order around the observed pair $(\pmb{\theta}_{i,\rm{obs}}, \pmb{\beta}_{i,\rm{obs}})$, where $\pmb{\beta}_{i,\rm{obs}}$ is again obtained by mapping the observed image position to the source plane,
	\begin{equation}
		\begin{aligned}
			\Delta t_{i,\rm{model}}
			& \approx \Delta t(\pmb{\theta}_{i,\rm{obs}}, \pmb{\beta}_{i,\rm{obs}}) \\
			& + (\pmb{\theta}_{i,\rm{model}}-\pmb{\theta}_{i,\rm{obs}}) \cdot \left[\nabla_{\pmb{\theta}} \Delta t\right]_{(\pmb{\theta}_{i,\rm{obs}}, \pmb{\beta}_{i,\rm{obs}})} \\
			& + (\pmb{\beta}_{\rm{model}}-\pmb{\beta}_{i,\rm{obs}}) \cdot \left[\nabla_{\pmb{\beta}} \Delta t\right]_{(\pmb{\theta}_{i,\rm{obs}}, \pmb{\beta}_{i,\rm{obs}})}.
		\end{aligned}
		\label{eq:td_expansion}
	\end{equation}
	By construction, $\pmb{\theta}_{i,\rm{obs}}$ is a stationary point of the arrival-time surface corresponding to the source position $\pmb{\beta}_{i,\rm{obs}}$ (Fermat's principle), so the image-plane gradient in \Eref{eq:td_expansion} vanishes identically, $\left[\nabla_{\pmb{\theta}}\Delta t\right]_{(\pmb{\theta}_{i,\rm{obs}}, \pmb{\beta}_{i,\rm{obs}})}=0$. Evaluating the surviving source-plane gradient from \Eref{eq:time_delay} then gives~\citep{2010PASJ...62.1017O},
	\begin{equation}
		\begin{aligned}
			\Delta t_{i,\rm{model}}
			& \approx \Delta t(\pmb{\theta}_{i,\rm{obs}}, \pmb{\beta}_{i,\rm{obs}}) + \frac{1+z_{\rm d}}{c}\frac{D_{\rm d}D_{\rm s}}{D_{\rm ds}}\,\theta_0^2 \, \times \\
			& (\pmb{\beta}_{\rm{model}}-\pmb{\beta}_{i,\rm{obs}}) \cdot (\pmb{\beta}_{i,\rm{obs}} - \pmb{\theta}_{i,\rm{obs}}),
		\end{aligned}
		\label{eq:td_approx}
	\end{equation}
	which, once again, involves only quantities evaluated at the observed image positions and the model source position.

	\subsection{Cluster galaxies}
	\label{ssec:scaling}
	In lensing by a galaxy cluster, one needs to account for the lensing from individual cluster galaxies, particularly those located in the vicinity of the lensed images. As the number of cluster galaxies is of the order of hundreds, modeling each cluster galaxy individually would lead to an under-constrained problem. Therefore, cluster galaxies are modeled using the pseudo-Jaffe ellipsoid~\citep[PJE;][]{2001astro.ph..2340K} mass distribution, with parameters determined by scaling relations between their observed luminosities and mass profile parameters, given by~\citep[e.g.,][]{2007NJPh....9..447J, 2011A&ARv..19...47K},
	\begin{equation}
		\begin{aligned}
			v_d           &= v_{d}^\star \left(\frac{L}{L^\star}\right)^\lambda,     \\
			\theta_{core} &= \theta_{core}^\star   \left(\frac{L}{L^\star}\right)^\beta, \\
			\theta_{cut}  &= \theta_{cut}^\star   \left(\frac{L}{L^\star}\right)^\alpha,
		\end{aligned}
		\label{eq:scaling}
	\end{equation}
	where~$L^\star$ is the typical luminosity of a galaxy at the cluster redshift and $v_{d}^\star$, $\theta_{core}^\star$, $\theta_{cut}^\star$ are its velocity dispersion, core radius, and truncation radius, respectively. With these scaling relations, modeling cluster galaxies introduces only a small set of free parameters, regardless of the number of cluster galaxies. In practice, the core radius of the $L^\star$ galaxy is poorly constrained by the lensing data alone and is therefore set to zero or some fixed value.
	
	\subsection{Implementation}
	\label{ssec:implementation}
	In \LensFactory, lens modeling is controlled by the \LensModel module, and the entire workflow, from model definition to posterior sampling, is driven by a single YAML configuration file. The configuration is organized into blocks specifying the observational setup, the background cosmology, the lens model, and the sampling settings. Any parameter of any lens component can be held fixed or assigned a uniform prior, including the source redshifts of image families lacking spectroscopic measurements. Unknown source redshifts can be sampled either directly in $z_s$ or through the corresponding distance ratio $a_{\rm dis}$. The latter is faster, since it avoids evaluating the cosmological distance integral at every step, but is valid only for single-plane lensing. The positional uncertainties can be circular or elliptical (Eq.~\ref{eq:chi2_img}), and the flux and time-delay terms enter the total $\chi^2$ only when the corresponding measurements are supplied. Both source-plane and image-plane minimization schemes are implemented. For the image plane, we implement the Newton-Raphson root-finding algorithm, which can be substantially faster once a good starting guess is available. In particular, we use the observed image positions themselves as natural initial guesses for the corresponding model images during minimization, avoiding the need to repeatedly search the full image plane from scratch at each step. However, this has the limitation that the lens model must be good, meaning that model images should lie near the observed images; otherwise, the algorithm can struggle. Hence, to use the above, one can first find a good fit with source-plane minimization and then start the image-plane minimization.
	
	For inference, \texttt{LensFactory.jl} currently provides the Nelder--Mead~(NM) downhill simplex optimizer~\citep{10.1093/comjnl/7.4.308} to locate the best-fit model, and two gradient-free samplers --- the Metropolis--Hastings (MH) algorithm and the affine-invariant ensemble sampler \citep[AIES;][]{2010CAMCS...5...65G} --- to explore the posterior around it. In a typical workflow, adopted throughout this work, a large number of independent NM runs, initialized randomly within the priors, are first performed to locate the global best-fit model; the AIES chains are then initialized in a small neighborhood of (i.e., jittered around) this solution to sample the posterior and estimate the parameter uncertainties. Independent optimizer runs, as well as individual walkers, are parallelized across threads using Julia's native multi-threading, which is what enables the full cluster-scale reconstructions presented in \Sref{sec:sims} and \Sref{sec:smacs} to complete within hours on a desktop-class machine. Owing to the modular structure of the package, alternative optimizers or samplers can be added with minimal effort (see also \Sref{sec:summary}). Worked examples covering the complete lens modeling workflow are available in the \Examples repository.

	\begin{figure*}[!h]
		\centering
		\includegraphics[width=1.0\linewidth]{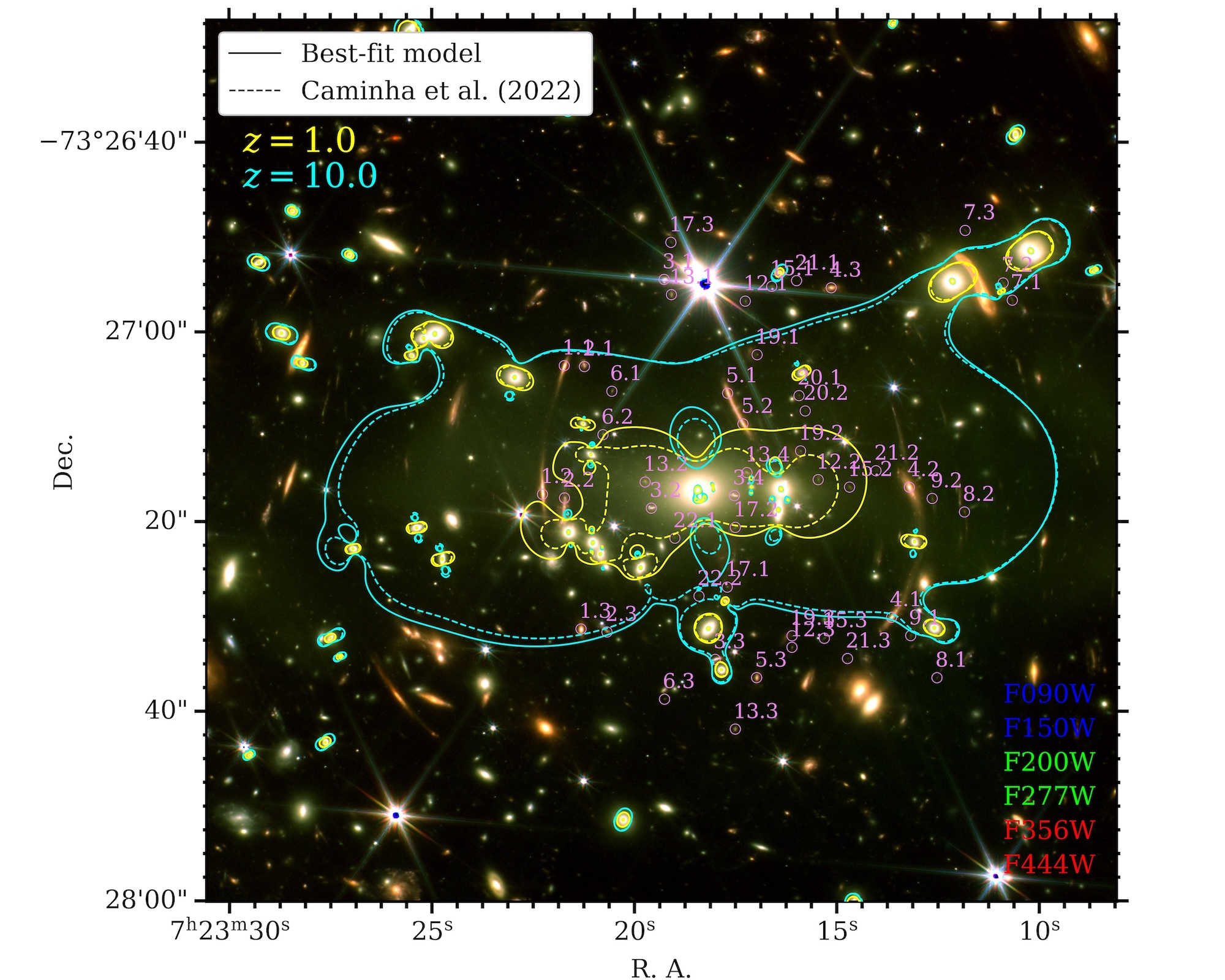}
		\caption{Color image of \smacs~($z=0.387$). The solid yellow and cyan curves represent the critical curves for $z_s=1$ and~$z_s=10$, respectively. For comparison, we also plot the critical curves corresponding to \citet{2022A&A...666L...9C}, shown by dashed curves.}
		\label{fig:smacs}
	\end{figure*}

	\begin{table*}
		\centering
		\caption{Best-fit parameter values corresponding to the \smacs strong lens model. Column~(1) shows the names of free parameters. Columns~(2) and~(3) represent the prior and best-fit values for each parameter. Column~(4) and~(5) represent the $1\sigma$~(i.e.,~[15.87, 84.13] percentile range) and $2\sigma$~(i.e.,~[2.28, 97.72] percentile range) error range around the best-fit value.}
		\label{tab:smacs}
		\begin{tabular}{lcccc}
			\hline
			Free parameters               & Prior         & Best-fit  & $1\sigma$  & $2\sigma$ \\
			\hline
			$\theta_{x,c}~['']$           & $[-2, 2]$     & $-0.41$   & $[-0.18, +0.68]$   & $[-0.58, +1.14]$    \\
			$\theta_{y,c}~['']$           & $[-2, 2]$     & $0.84$    & $[-0.06, +0.16]$   & $[-0.17, +0.27]$    \\
			$v_d~[\rm{km/s}]$             & $[500, 1500]$ & $1313.69$ & $[-24.56, +53.92]$ & $[-65.67, +95.43]$  \\
			$\theta_s~['']$               & $[0.1, 30]$   & $16.38$   & $[-0.61, +1.66]$   & $[-1.75, +2.90]$    \\
			$\epsilon$                    & $[0.1, 0.7]$  & $0.23$    & $[-0.01, +0.07]$   & $[-0.05, +0.11]$    \\
			$\phi_{\rm PA}~[\rm{deg}]$    & $[-45, 45]$   & $0.35$    & $[-0.67, +1.93]$   & $[-2.61, +2.96]$    \\
			$\gamma_{\rm ext}$            & $[0.0, 0.2]$  & $0.07$    & $[-0.02, +0.02]$   & $[-0.03, +0.06]$    \\
			$\phi_{\gamma}~[\rm{deg}]$    & $[-90, 90]$   & $-57.30$  & $[-5.47, +33.18]$  & $[-13.59, +47.71]$  \\
			$v_{d}^\star~[\rm{km/s}]$     & $[200, 500]$  & $475.43$  & $[-123.22, +0.46]$ & $[-180.72, +21.09]$ \\
			$\theta^\star_{\rm cut}~['']$ & $[0.1, 30]$   & $1.32$    & $[-0.02, +1.24]$   & $[-0.22, +2.70]$    \\
			\hline
		\end{tabular}
	\end{table*}
	
	\section{Simulated galaxy clusters}
	\label{sec:sims}
	We start by using the \LensFactory mass reconstruction pipeline on two simulated galaxy clusters from \citet[hereafter~\citetalias{2017MNRAS.472.3177M}]{2017MNRAS.472.3177M}\footnote{\url{http://pico.oabo.inaf.it/~massimo/Public/FF/index.html}}, namely, Ares~($z=0.5$; \Fref{fig:ares}) and Hera~($z=0.507$; \Fref{fig:hera}). \citetalias{2017MNRAS.472.3177M} presented a cluster mass modeling challenge where, in short, observations of strong lensing galaxy clusters by HST were simulated, and mock catalogs were given to multiple teams of lens modelers to reconstruct the cluster lens mass distribution. Each team performed strong lens modeling in a blind manner, and the reconstructed mass models were compared with the true mass model for a range of properties (such as mass profile, magnification, convergence, ellipticity, orientation). We note that our mass reconstruction of Ares and Hera is not blind and cannot be evaluated on an equal footing as the mass reconstruction presented in \citetalias{2017MNRAS.472.3177M}. However, these simulated clusters can still be used to validate both the accuracy and the efficiency of the overall lens modeling pipeline. Below, we describe our lens modeling of these clusters and compare the resulting models to the truth. The input YAML files for lens modeling are publicly available in the \Examples repository. 
	
	\subsection{Ares}
	\label{ssec:ares}
	Ares was simulated using parametric density profiles for various mass components with~$H_0=70.4~\rm{km/s/Mpc}$ and~$(\Omega_{m,0}, \Omega_{r,0}, \Omega_{w,0}) = (0.272, 0.0, 0.728)$. In our current work, we fix the cosmological parameters during lens mass reconstruction to the same values used to generate the simulations.
	
	For our lens reconstruction, we use the publicly available lensed image and cluster galaxy catalogs. In total, we have 85 sources leading to 242 lensed images and 331 cluster galaxies, with source redshifts fixed to their catalog values. We place two dark matter halos, modeled using elliptical NFW profiles, at the positions of two of the brightest cluster galaxies. The NFW profiles were approximated using the method described in \citet{2021PASP..133g4504O}. Each of these two halos introduces four free parameters~(i.e., mass, concentration, ellipticity, and PA) as their positions are fixed. The cluster galaxies are modeled using PJE profiles assuming axial symmetry and~$\theta_{\rm core}^\star=0$. We also fix the power-law indices for the scaling relation to~$(\lambda, \beta, \alpha)=(0.25, 0.50, 0.50)$. The reference magnitude for scaling relations is fixed to~$m_{\star}=18.5$ in the F814W band. Three bright galaxies, marked by red-dashed circles in \Fref{fig:ares}, are modeled separately using PJE profiles. Each of these galaxies also introduces four free parameters~(i.e., velocity dispersion, ellipticity, PA, and cut radius; core radius is again fixed to zero), with their positions fixed. With the above set-up, the total number of free parameters is~22 (i.e., 8 for the two dark matter halos, 12 for the three galaxies, and 2 for the scaling relation).
	
	To find the best-fit model, we employ a two-step process. First, we perform~$10^4$ NM~optimizer runs and find the best-fit among these. In the next step, we take the above best-fit and run the AIES with 32 chains each with $10^5$ steps. In both of the above steps, the~$\chi^2$ was calculated in the source plane, assuming a (circular) positional uncertainty of $\sigma = 0.5''$ per image. The model was run on an M3~Ultra machine while utilizing 16 cores. The total wall-clock runtime for Ares was $\simeq11$~hours. The final best-fit lens model has an image-plane RMS of~$0.44''$.
	
	Since our reconstruction is not blind (knowledge of the true mass distribution could, in principle, inform our modeling choices), a detailed quantitative comparison with the truth would carry limited weight. We therefore restrict ourselves to a qualitative comparison of the critical curves, convergence, and magnification against the true maps. A comparison of critical curves, for multiple source redshifts, corresponding to true and reconstructed mass distributions for Ares is shown in \Fref{fig:ares}. We find that at all redshifts, the best-fit and true critical curves closely match, indicating high fidelity of the reconstructed mass distribution. Looking at the relative difference in convergence (left panel in \Fref{fig:dkappa}), we again note that the overall distribution of surface density is reproduced well. However, there are also circular artifacts. These artifacts arise because, during the Ares simulation, the PIEMD profiles corresponding to galaxy components were sharply truncated at~$\theta=\theta_{\rm cut}$, as highlighted in~\citetalias{2017MNRAS.472.3177M}, whereas the galaxy component profile used during our mass reconstruction has no such sharp truncation. The relative difference in magnification for Ares is shown in the left panel of \Fref{fig:dmu}.

	\subsection{Hera}
	\label{ssec:hera}
	Hera was directly derived from an N-body simulation with~$H_0=72~\rm{km/s/Mpc}$ and~$(\Omega_{m,0}, \Omega_{r,0}, \Omega_{w,0}) = (0.24, 0.0, 0.76)$. Unlike Ares, the underlying mass distribution of Hera is not built from analytic profiles, so a parametric lens model is expected to only approximate its true structure. This makes Hera a more realistic and more challenging test of the mass reconstruction pipeline.
	
	For our lens reconstruction, we again use the publicly available lensed image and cluster galaxy catalogs. In total, we use all 19~sources leading to 65~lensed images and 340~cluster galaxies, with source redshifts fixed to their catalog values. We place two dark matter halos, modeled using (approximated) elliptical NFW profiles, at the positions of the two brightest cluster galaxies, with their positions held fixed. Each halo introduces four free parameters~(i.e., mass, concentration, ellipticity, and position angle). In addition, we include external shear along with a third-order~($m=3$) multipole component with radial dependence $\propto \theta^{2}$, to capture the angular structure of the mass distribution. Both introduce two additional free parameters each~(i.e., amplitude and orientation). The cluster galaxies are modeled using PJE profiles assuming axial symmetry and~$\theta_{\rm core}^\star=0$, with the power-law indices of the scaling relation fixed to~$(\lambda, \beta, \alpha)=(0.25, 0.50, 0.50)$ and the reference magnitude fixed to~$m_{\star}=19.83$ in the F814W band. With the above set-up, the total number of free parameters is~14 (i.e., 8~for the dark matter halos, 2~for external shear, 2~for the multipole, and 2~for the scaling relation).
	
	To find the best-fit model, we follow the same two-step procedure as for Ares, with the~$\chi^2$ again calculated in the source plane assuming a (circular) positional uncertainty of~$\sigma=0.5''$ per image. Here, we run the AIES with 48~chains, each with $10^5$~steps. The total wall-clock runtime for Hera was $\simeq1.3$~hours on the same machine. The final best-fit lens model has an image-plane RMS of~$0.49''$.
	
	A comparison of the true and reconstructed critical curves at multiple source redshifts is shown in \Fref{fig:hera} and is comparable to the best model. Compared to Ares, we notice larger deviations in the positions of the critical curves at all three redshifts. The most significant differences occur at $z=9$ around the two bright galaxies in the upper-left and bottom-right quadrants of the image. Looking at the multiple image distribution in Fig.~4 of \citetalias{2017MNRAS.472.3177M}, we note that both of these galaxies lie outside the region covered by the lensed images, and hence the mass distribution around them is essentially unconstrained by the lensing data. The same can also be seen in the relative difference maps for convergence and magnification, shown in the right panels of \Fref{fig:dkappa} and \Fref{fig:dmu}, respectively, where the largest residuals lie outside the multiply imaged region. Within the region constrained by the lensed images, the convergence is recovered reasonably well, although the residuals are larger than for Hera, reflecting the substructure and asymmetry of the N-body mass distribution that the two-halo parametric model cannot fully capture.
	
Although our reconstructions for Ares and Hera are not blind and therefore cannot be placed on an equal footing with the models submitted to \citetalias{2017MNRAS.472.3177M}, it is still instructive to compare them with the true mass models. The image-plane RMS values obtained here, $0.44''$ for Ares and $0.49''$ for Hera, are comparable to those of the best-performing models in the challenge: for Ares, M17 report RMS values ranging from $\simeq0.27''$ (Glafic) to $1.8''$ (for Zitrin-LTM), and for Hera from $0.43''$ (Glafic) to $1.2''$ (for Zitrin-LTM). The same pattern is also seen for convergence and magnification maps.

	\begin{figure}[!h]
		\centering
		\includegraphics[width=1.0\linewidth]{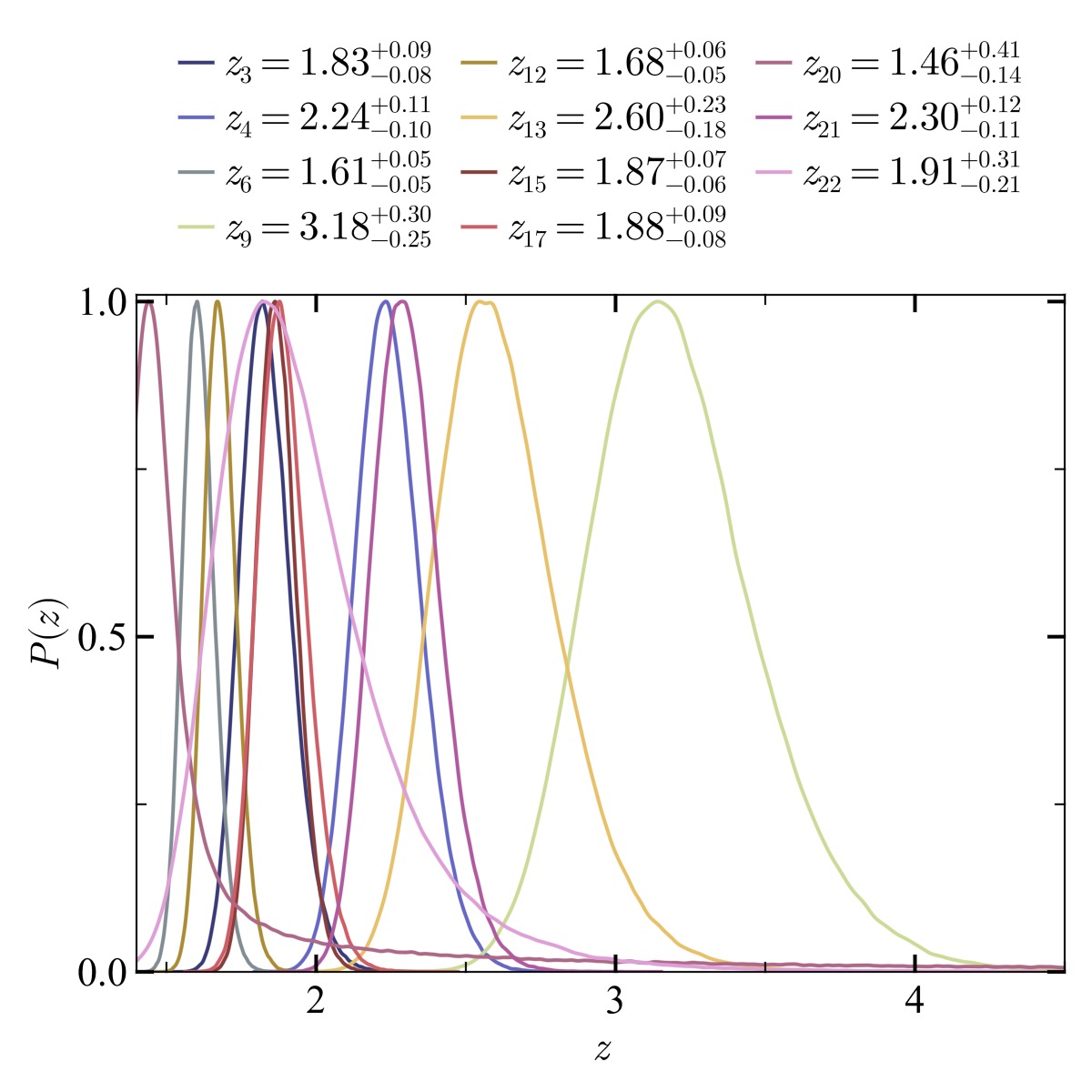}
		\caption{Probability density (normalized) for lens model free redshifts corresponding to various multiple image families. The median along with~$1\sigma$ errors~(i.e., 16th and 84th percentile range) are shown at the top of the plot. Here, we do not show system~8 because the corresponding redshift is not well constrained~($z>10$).}
		\label{fig:smacs_z}
	\end{figure}

	\section{SMACS~J0723.3$-$7327}
	\label{sec:smacs}
	SMACS~J0723.3$-$7327 (hereafter SMACS~J0723; \citealt{2018MNRAS.479..844R}) was the first strong-lensing galaxy cluster targeted by JWST as part of the Early Release Observations~(ERO; \citealt{2022ApJ...936L..14P}). Before JWST, it was observed with HST under the Reionization Lensing Cluster Survey~(RELICS; \citealt{2019ApJ...884...85C}) treasury program\footnote{\url{https://archive.stsci.edu/hlsp/relics}}\textsuperscript{,}\footnote{SMACS~J0723 has also been observed with HST under the GO-11103, GO-12166, GO-12884, and GO-16729 programs.}, and with the Multi-Unit Spectroscopic Explorer~(MUSE; \citealt{2010SPIE.7735E..08B}) at the Very Large Telescope.
	
	Multiple strong lens models for SMACS~J0723 have been presented in recent years. \citet{2022ApJ...938...14G} presented an HST-based strong lens model with an RMS of~$1.7''$ utilizing five multiple image families. With JWST observations, \citet{2022A&A...666L...9C}, \citet{2022ApJ...938L...6P}, and \citet{2023ApJ...945...49M} presented updated lens models achieving RMS values of $0.51''$, $0.48''$, and $0.32''$, respectively, though we note that these are based on somewhat different image samples and model complexities, and hence are not directly comparable. In particular, the model of \citet{2023ApJ...945...49M} has the most number of free parameters, whereas that of \citet{2022A&A...666L...9C} has the least.

	\begin{figure*}[!h]
		\centering
		\includegraphics[width=0.48\linewidth]{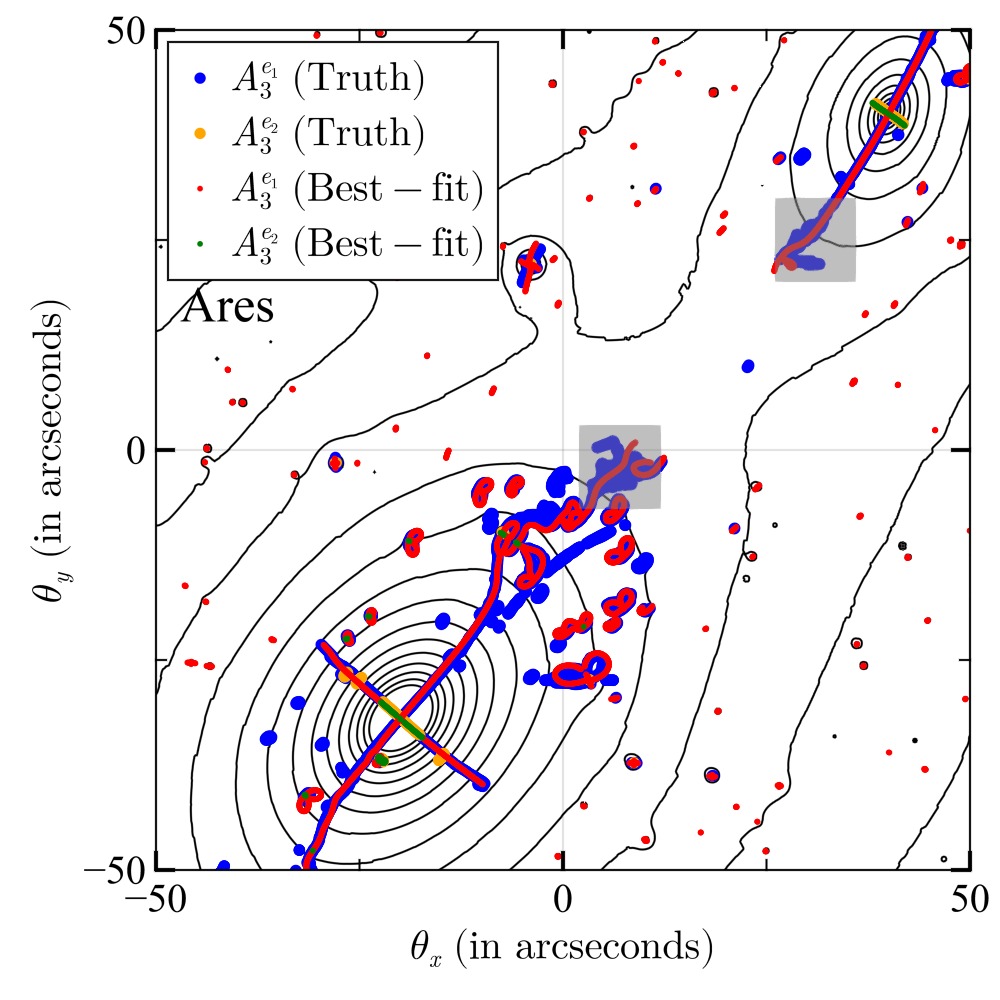}
		\includegraphics[width=0.48\linewidth]{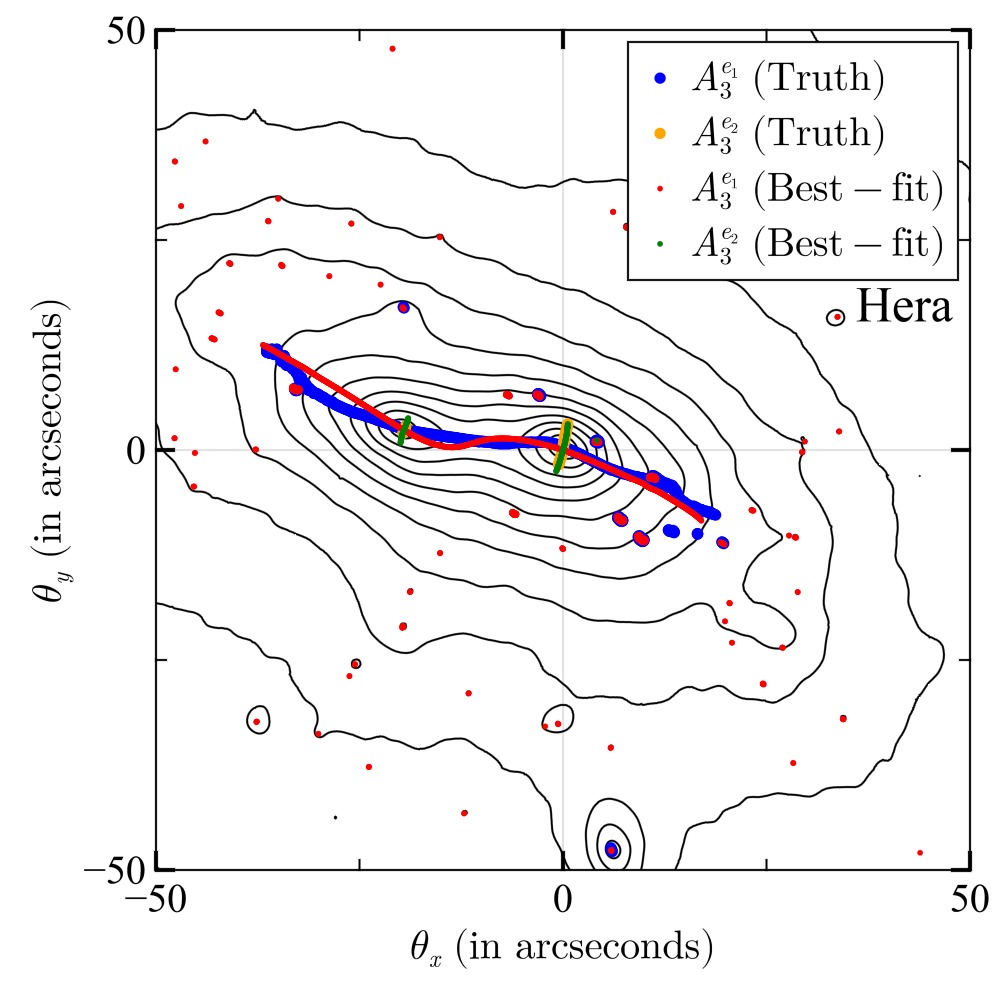}
		\caption{Comparison of singularity maps~(true vs. best-fit) for Ares and Hera in left and right panels, respectively. The black contours in both panels represent the constant convergence~($\kappa$) for~$z_s=9$. Blue and yellow points represent~$A_3^{e_1}$- and~$A_3^{e_2}$-lines corresponding to the true mass distribution. Red and green points show~$A_3^{e_1}$- and~$A_3^{e_2}$-lines obtained using the best-fit lens model. The $A_3^{e_1}$- and~$A_3^{e_2}$-lines correspond to the~$e_1$~(i.e., tangential) and $e_2$~(i.e., radial) eigenvalues of the deformation tensor. Examples of spurious $A_3$-lines for Ares are marked by gray boxes.}
		\label{fig:singularity}
	\end{figure*}
	
	Here, following \citet{2022A&A...666L...9C}, we construct a strong-lensing mass model using the same catalog of cluster galaxies and multiple image families, consisting of 17~multiple image families leading to 49~lensed images, together with 62~cluster galaxies. Of these families, 5~have spectroscopic redshifts, which we hold fixed; for the remaining 12~systems, we leave the redshifts free. In our model, the cluster-scale dark matter halo is modeled using an isothermal ellipsoid with a finite core radius~(implemented as \texttt{SIELens} in \LensFactory), cluster galaxies are modeled as PJE profiles, along with external shear\footnote{\citet{2022A&A...666L...9C} highlighted that the model with a single cluster-scale halo is a better fit based on the Akaike information criterion~\citep[AIC;][]{1974ITAC...19..716A} and Bayesian information criterion~\citep[BIC;][]{1978AnSta...6..461S}.}. The dark matter halo has six free parameters: its position~($\theta_x$, $\theta_y$), velocity dispersion~($v_d$), core radius~($\theta_s$), ellipticity~($\epsilon$), and position angle~($\phi_{\rm PA}$). All cluster galaxies are modeled using the scaling relations (i.e., Eq.~\ref{eq:scaling}) with $(\lambda, \beta, \alpha)=(0.25, 0.50, 0.50)$. The reference magnitude and core radius are fixed to $m_{\star}=17.66$ in the F160W~band and $\theta_{\rm core}^\star=0.05''$, respectively. With that, the total number of free parameters related to the lens model is ten~(i.e., six for the dark matter halo, two for the external shear, and two for the scaling relations, namely $v_{d,\star}$ and $\theta_{{\rm cut},\star}$), in addition to the 12~free source redshifts. We sample uniformly in the distance ratio~$a_{\rm dis}$~(\Sref{sec:basics}) rather than in redshift directly.  We note that the choice of a cored isothermal ellipsoid here, rather than the elliptical NFW profile used for Ares and Hera, is deliberate. Since one of our aims is to test the consistency of independently developed modeling codes. Hence, we decided to use the above parameterization for the cluster-scale halo, so that any significant differences with respect to their model can be attributed to the modeling code and inference procedure rather than to differing model assumptions.
	
	To construct the lens model, we again follow the two-step procedure with source-plane minimization, assuming a (circular) positional uncertainty of $\sigma=0.5''$ per image. First, we perform $10^5$~runs of the NM optimizer, and then we run 48~AIES chains with $10^6$~steps jittered around the NM best-fit solution. The total wall-clock runtime is $\simeq4$~hours on the same machine as above. The resulting best-fit model has an image-plane RMS of~$0.46''$, lower than the value of $0.51''$ reported by \citet{2022A&A...666L...9C} for the same set of constraints.
	
	The best-fit lens parameters, along with their uncertainties, are listed in Table~\ref{tab:smacs}. We note that the best-fit value for~$v_d^\star$ is close to the upper bound, where the best-fit value for~$\theta_{\rm cut}^\star$ prefers smaller values. This highlights a degeneracy between these two parameters. To check this, we ran another model widening the prior on~$v_d^\star$ upto 600~km/s, and we still get the same behavior.
	
	A comparison of the critical curves corresponding to our best-fit model and that of \citet{2022A&A...666L...9C} is shown in \Fref{fig:smacs}. Overall, the two sets of critical curves match very well, particularly at~$z_s=10$. At $z_s=1$, however, they show visible differences; this is expected, since most of the multiple-image families lie at higher redshifts, so the lensing data primarily constrain the deflection field at large distance ratios, whereas the $z_s=1$ critical curve lies deep in the cluster core, where its location is sensitive to the (poorly constrained) inner mass profile. The inferred redshift distributions for the various image families are shown in \Fref{fig:smacs_z}, and are consistent with the model redshifts reported by~\citet{2022A&A...666L...9C}. For system~8, the posterior remains unconstrained at the upper end of the prior range, $z\in[4,15]$; the model favors $z>10$, again consistent with \citet{2022A&A...666L...9C}, who similarly found the redshift of this system to be unconstrained. Spectroscopic confirmation will likely be required to determine the redshift of this system. 
	
	To further assess the reliability of our mass reconstruction, we compare the projected mass enclosed within several apertures against values from the literature. From our best-fit model, we find 
	$M(<78\,\rm{kpc})=4.17\times10^{13}\,\mathrm{M_\odot}$, 
	$M(<90\,\rm{kpc})=5.24\times10^{13}\,\mathrm{M_\odot}$, and 
	$M(<128\,\rm{kpc})=8.91\times10^{13}\,\mathrm{M_\odot}$. These are in good agreement with previously published strong-lensing models~\citep{2022A&A...666L...9C, 2023ApJS..264...15S}. The agreement in enclosed mass across independently constructed models --- despite differences in modeling code, profile parameterization, and best-fit parameters --- illustrates the robustness of the recovered mass distribution in the region constrained by the lensed images.
	
	The input YAML configuration, the multiple-image and cluster-galaxy catalogs, and the posterior samples corresponding to our best-fit model are publicly available in the \texttt{ClusterLensModels} repository\footnote{\url{https://github.com/akmeena766/ClusterLensModels}}, which we intend to maintain as a growing collection of cluster lens models constructed with \LensFactory.
	
	\section{Singularity maps}
	\label{sec:singularity}
	In strong lensing, the standard observables --- lensed image positions, their time delays, and flux ratios --- constrain the lensing potential and its first- and second-order derivatives at the image positions. The higher-order derivatives, however, remain largely unconstrained by these observables. Therefore, a follow-up question is whether higher-order derivatives also agree across lens models. A natural diagnostic for comparing higher-order derivatives across lens models is to compare their corresponding singularity maps~\citep{2020MNRAS.492.3294M, 2021MNRAS.503.2097M, 2021MNRAS.506.1526M}. 
	
	A singularity refers to the points where the determinant of the Jacobian of the lens mapping vanishes and can be divided into two types: stable and unstable. The stable singularities of lens mapping are folds and cusps, and occur for all lens models. Unstable~(point) singularities, such as swallowtails~($A_4$) and umbilics~($D_4$), only occur at specific source redshifts for a given lens model. A singularity map consists of $A_3$-lines and point singularities. The $A_3$-lines form the backbone of the singularity map and trace the location of cusp points (which form in the source plane at all redshifts) in the image plane. Since cusp forms on both tangential and radial caustics, we have two corresponding~$A_3$lines. The unstable~(point) singularities of the lens mapping occur only for specific source redshifts, and a small perturbation in lens system parameters can make them vanish~\citep{2020MNRAS.492.3294M}. A source lying close to these leads to characteristic (exotic) image formations. Since $A_3$-lines and point singularities depend on the third- and higher-order derivatives of the lensing potential, the singularity map provides a direct, albeit restricted, probe of the higher-order derivative structure of the lens model --- constrained to specific locations and specific combinations of derivatives dictated by the catastrophe conditions. We refer the reader to Chapter~6 in \citet{1992grle.book.....S} for singularities in gravitational lensing, and for our singularity map formalism to~\citet{2020MNRAS.492.3294M}.
	
	In \LensFactory, we have implemented a module, named \SingularityMap, that calculates the singularity map for a given lens model. Since the lens models reconstructed above are built from analytic profiles, the lensing quantities can be sampled at arbitrarily high resolution, leading to very high-precision singularity maps. This is not the case for the publicly available convergence and shear maps of Ares and Hera, where finite resolution~(combined with the 32-bit floating point precision) limits the precision of higher-order derivatives, introducing spurious features. We therefore restrict the qualitative comparison to $A_3$-lines in the central regions of these clusters with~$a_{\rm dis}\leq0.5$~(i.e.,~$z_s\leq1.2$), which is shown in \Fref{fig:singularity}. We note that the singularity map formalism, as implemented here, assumes single-plane lensing.
	
	For Ares (left panel), we find that the overall spine of the best-fit model $A_3$-lines close to the two dark matter halos follows the truth very well. Even the cusp transition points between~$A_3^{e_1}$- and~$A_3^{e_2}$-lines lie very close in the image plane. However, as we move to the region in between the two halos~(close to the center of the panel), we start to see considerable differences in $A_3$-lines. These differences arise from two distinct sources: numerical artifacts in the publicly available maps — examples of which are highlighted by gray boxes in \Fref{fig:singularity} — and genuine structures in the true mass distribution that are not captured by the best-fit model. The latter likely reflects the limited constraining power of the available multiple image positions in the inter-halo region, where the higher-order derivatives of the lensing potential are insufficiently constrained by the data. For Hera~(right panel), we see that the best-fit $A_3$-line spine deviates from the truth as we move farther from the center. Even though the lens model includes a third-order multipole component to capture angular structure in the mass distribution, the deviation likely reflects the radial variation of the position angle twist in the actual mass distribution, which is continuous and may not be fully captured by the assumed radial dependence of the multipole component of the lens model, particularly in regions with sparse observational constraints. A quantitative analysis would be needed to disentangle the contributions of model flexibility and data insufficiency, which we defer to future work.
	
	The comparisons above suggest that singularity maps provide a sensitive diagnostic for evaluating lens models beyond the standard observables, probing the higher-order derivative structure of the lensing potential in a way that convergence map comparisons and image position residuals alone cannot. A quantitative assessment of the connection between $A_3$-line agreement and observational constraint density is beyond the scope of this work. We therefore propose the inclusion of singularity map comparisons as a complementary diagnostic in future lens modeling challenges, with the longer-term goal of using observed exotic image configurations to directly constrain the higher-order derivative structure of cluster lenses.

	\section{Summary and future prospects}
	\label{sec:summary}
	In this work, we described \LensFactory, an efficient, open-source, general-purpose strong lens modeling package. The package supports both single- and multi-plane lensing along with a wide range of analytic lens profiles, with a modular structure (built using Julia's multiple dispatch) that makes adding new profiles, inference methods, and functionalities straightforward. \LensFactory implements all standard lensing quantities, numerically solves the lens equation to predict image positions, and constrains lens mass models through a combination of optimization and posterior sampling, with computationally expensive steps parallelized via \texttt{Julia}'s native multi-threading. The underlying implementation is low-level and type-stable, with performance-critical routines operating in place on preallocated arrays, enabling full~(typical) cluster-scale lens reconstructions on timescales of hours rather than days on a desktop-class machine.
	
	We validated the lens mass reconstruction part of the package using two simulated galaxy clusters, Ares and Hera from \citetalias{2017MNRAS.472.3177M}. For both clusters, the reconstructed critical curves closely follow those of the true mass distributions, and the relative differences in convergence and magnification maps show good overall agreement. We further applied \LensFactory to the JWST-observed cluster \smacs, constructing a strong lens model based on the cluster-galaxy catalog and multiple-image families provided in~\citet{2022A&A...666L...9C}, and obtained results consistent with previously published models, again validating our pipeline.
	
	In addition, we proposed singularity maps as a complementary diagnostic for comparing lens models. Since the A3-lines and point singularities depend on third- and higher-order derivatives of the lensing potential, singularity maps probe the derivative structure of lens models beyond what standard observables (image positions, time delays, and flux ratios) constrain. Applying this diagnostic to Ares and Hera, we found that the best-fit $A_3$-lines fit Ares well but deviate for Hera, whose mass distribution is more complex, suggesting that singularity-map comparisons can serve as a sensitive diagnostic in future lens modeling challenges.
	
	There remains considerable scope to add new functionality and improve the existing implementation. The immediate developments we plan for the near future are as follows:
	\begin{enumerate}    
		\item Currently, \LensFactory ships with MH and AIES samplers, both of which are gradient-free. The choice of these samplers is motivated by their simplicity and easy implementation. Gradient-based samplers such as Hamiltonian Monte Carlo~(HMC) or the No-U-Turn Sampler~(NUTS) can substantially improve efficiency in high-dimensions~(needed for complex lenses with a large number of free parameters). We plan to include support for gradient-based samplers via \texttt{Turing.jl}\footnote{\url{https://turinglang.org/}}~\citep{10.1145/3711897} in the near future.
		
		\item In singularity map construction, currently, the resolution is fixed. However, since our lens model is made of analytic profiles, we can calculate lensing quantities on an arbitrarily high-resolution grid. This opens the possibility of adaptively refining the singularity maps to accurately determine the redshifts of various point singularities and reduce noise. Hence, we plan to update the \SingularityMap module to include adaptive refinement to improve the resulting singularity maps.
		
		\item Recently, attempts have also been made to model the full surface brightness of giant arcs in cluster lenses~\citep[e.g.,][]{2024ApJ...976..110A, 2026ApJ..1003..146E}. At present, \LensFactory does not have the capability to model extended sources, but we expect to add this capability in the future.
	\end{enumerate}

	\section*{Acknowledgements}
	The author thanks Wenlei Chen for providing reduced NIRCam data products of \smacs. The author thanks reviewer for useful comments. The author acknowledges the support from the Start-up Grant IE/CARE-25-0305 provided by the IISc, Bengaluru, India. This research has made use of NASA’s Astrophysics Data System Bibliographic Services. 
	
	This work utilizes the following packages: \href{https://julialang.org/}{\texttt{Julia}}~\citep{Julia2017}, \href{https://akmeena766.github.io/LensFactory.jl/stable/}{\texttt{LensFactory.jl}}, \href{https://makie.org/website/}{\texttt{Makie.jl}}~\citep{2021JOSS....6.3349D}.

	\bibliography{Reference}
	
\end{document}